\documentclass[journal=jacsat,manuscript=article,layout=twocolumn]{achemso}

\usepackage[version=3]{mhchem} 
\usepackage{nicefrac}
\usepackage{amssymb}
\usepackage{lettrine}
\usepackage{color,soul}
\usepackage{newtxtext}
\usepackage[scaled=0.9]{helvet}
\usepackage[fontsize=9pt]{fontsize}
\usepackage[version = 3]{mhchem}
\usepackage{graphicx}
\usepackage{amsmath}
\usepackage{amsfonts}
\usepackage{chemformula}
\usepackage[flushleft]{threeparttable}
\usepackage{multirow}
\usepackage{array}
\usepackage{ascii}
\usepackage[T1]{fontenc}
\usepackage{float}
\usepackage{placeins}

\author{Nikhil S. Chellam}
\affiliation[Northwestern University]
{Department of Chemical \& Biological Engineering, Northwestern University, Evanston, IL 60208}
\alsoaffiliation[Northwestern University]
{International Institute for Nanotechnology, Northwestern University, Evanston, IL 60208}
\author{George C. Schatz}
\affiliation[Northwestern University]
{Department of Chemical \& Biological Engineering, Northwestern University, Evanston, IL 60208}
\alsoaffiliation[Northwestern University]
{Department of Chemistry, Northwestern University, Evanston, IL 60208}
\alsoaffiliation[Northwestern University]
{International Institute for Nanotechnology, Northwestern University, Evanston, IL 60208}
\email{g-schatz@northwestern.edu}

\title[Manuscript Title]
  {Density Functional Tight-Binding Captures Plasmon-Driven H$_2$ Dissociation on Al Nanocrystals}

\abbreviations{IR,NMR,UV}

\begin{document}
\abstract{Aluminum nanocrystals offer a promising platform for plasmonic photocatalysis, yet a detailed understanding of their electron dynamics and consequent photocatalytic performance has been challenging thus far due to computational limitations. Here, we employ density functional tight-binding methods (DFTB) to investigate the optical properties and \ch{H2} dissociation dynamics of Al nanocrystals with varying sizes and geometries. Our real-time simulations reveal that Al's unique free-electron nature enables efficient light-matter interactions and rapid electronic thermalization. Cubic and octahedral nanocrystals ranging from 0.5 to 4.5 nm exhibit size-dependent plasmon resonances in the UV, with distinct spectral features arising from the particle geometry and electronic structure. By simulating \ch{H2} dissociation near Al nanocrystals, we demonstrate that hot electrons generated through plasmon excitation can overcome the molecule's strong chemical bond within tens of femtoseconds. The laser intensity threshold is comparable to previous reports for Ag nanocrystals, though significantly lower than that of Au. Notably, the dipolar plasmon mode demonstrates higher efficiency for this reaction than the localized interband transition for particles at these sizes. Taken together, this work provides mechanistic insights into plasmon-driven catalysis and showcases DFTB's capability to study quantum plasmonics at unprecedented length and time scales.}
\section{Introduction}
\lettrine{A}{luminum} can be considered the archetypal free electron metal. Indeed, with three electrons available in its conduction band, \ch{Al}'s high free electron density endows it with a high plasma frequency, enabling strong light-matter interactions throughout the optical spectrum.\cite{Chan_2008, Gerard_2015} At the nanoscale, spatial confinement of its collective plasma oscillations renders its localized surface plasmon resonances (LSPRs) tunable from the deep UV, through the visible, and beyond the IR.\cite{Knight_2014} With increasing sizes, its polarizability is sufficiently high such that even a 30 nm spherical (i.e., isotropic) Al nanocrystal is predicted to sustain higher-order multipolar modes,\cite{Ross_2015} a characteristic its coinage metal counterparts do not possess unless much larger (beyond 150 nm) or with a suitably low particle symmetry.\cite{Kelly_2003, Millstone_2005} And, perhaps most attractively, Al is the most abundant metal in the Earth's crust and has a high CMOS compatibility; recent developments in colloidal and lithographic syntheses of Al nanocrystals,\cite{McClain_2015, Knight_2014} while nonetheless an emerging field of interest, may allow their application in a wide variety of Earth-abundant plasmon-driven catalytic and metamaterial systems.\cite{Juarez_2024, Freire-Fernandez_2023, Swearer_2016, Zhou_2016}

\indent The plasmon itself is a femtosecond-scale phenomenon. Once an LSPR mode has been excited, within the first few femtoseconds, the dominant process is dephasing to electron-hole pairs with energies equivalent to that of the LSPR.\cite{Brown_2016, Stefancu_2024} Following plasmon dephasing, these electronically "hot" carriers will redistribute their energies through additional scattering events with photons (leading to luminescence), as well as other low-energy electrons from 100 fs to 1 ps. The reduced energy of these scattered electrons increases their propensity to scatter with phonons over several ps leading to thermal dissipation from 100 ps to 10 ns.\cite{Brongersma_2015} In turn, these scattering processes can raise the electrons' translational temperatures to thousands of degrees higher than that of their environment, providing a highly efficient means by which electrons can transfer into nearby adsorbates\cite{Fojt_2022, Boerigter_2016, Vanzan_2023} and thereby confer enough energy to induce otherwise energetically unfavorable chemical reactions under ambient conditions.\cite{Christopher_2011, Linic_2015, Mukherjee_2013, Swearer_2016}

\indent Other mechanisms have been proposed to describe plasmon-driven catalysis beyond the hot carrier picture. A related but distinct process involves direct electronic excitation between metal-adsorbate states,\cite{Kazuma_2020} which requires the adsorbate to be close to the nanoparticle surface and have a compatible excitation energy to that of the charge transfer mode. Others have proposed that even the LSPR electric field enhancement or localized heating\cite{Zhou_2018, Seemala_2019} are themselves capable of inducing excitation and dissociation events (\emph{via} vibronic excitation) without direct adsorption onto and subsequent charge transfer from the metal surface. Despite debates regarding their underlying mechanisms, plasmon-driven photocatalysis has been experimentally demonstrated to induce dissociation of many industrially-relevant molecules, including \ch{H2},\cite{Mukherjee_2013, Wu2020} \ch{H2O},\cite{Lee_2012} \ch{CH3OH},\cite{Bayles_2022} \ch{H2S},\cite{Lou_2022} and \ch{NH3},\cite{Yuan_2022} which further underscores its promise as a sustainable alternative to their thermocatalytic counterparts.

\indent From a theoretical standpoint, first-principles modeling and disentanglement of these processes is challenging due to the large number of electrons and electronic excited states involved. This is particularly the case for metals (and even more so for free-electron metals), where, due to their continuous density of states, electronic convergence is difficult, if not impossible, for systems sufficiently large to even be considered plasmonic.\cite{Aikens_2008, Zhou2016, Seveur_2023} Moreover, the tens to hundreds of fs timescales involved in plasmon-driven photodissociation effectively lead to a balance in such \textit{ab initio} studies between using small clusters\cite{Bao_2019, Wu2020, Chen_2023} and short pulses limited to the first tens of fs.\cite{Zhang_2018, Li_2023}

\indent While conceptualized nearly four decades ago,\cite{Seifert_1986, Elstner_1998} density functional tight-binding (DFTB) methods have experienced a renaissance and can, in fact, be used to describe these photochemical processes using suitably large nanocrystals with timescales extending into the picosecond regime. Indeed, herein we utilize this semiempirical method to study the electron dynamics of \ch{H2} dissociation in the proximity of bare (i.e., unoxidized) Al nanoclusters. Perhaps counterintuitively, we find that such a tight-binding method can reasonably describe the optical properties of this free-electron metal, with strong deviations based on particle symmetry due to their ability to sustain different plasmon modes. Importantly, we show that the laser intensity threshold for \ch{H2} dissociation is comparable to Ag while significantly lower than Au,\cite{Giri_2023} with the dipolar plasmon mode having higher efficiency than the interband transition at these sizes.

\section{Theoretical Methods}
All calculations in this work were performed using the open-source DFTB+ v. 22.2 software package.\cite{Hourahine_2020} Herein, we utilized the publicly-available {\asciifamily matsci-0-3} parameter set to simulate Al and its interactions with \ch{H2}.\cite{Frenzel_2005, Guimaraes_2007} Together, this method provides a balance between accuracy and computational cost, rendering it suitable for larger-scale simulations.

Octahedral and cubic \ch{Al} nanocrystals were constructed by truncating a bulk Al unit cell along the $\langle 111 \rangle$ and $\langle 100 \rangle$ directions, respectively. Octahedral nanocrystals were constructed using $n = \{19, 44, 85, 146, 231, 344, 489, 670, 891\}$ atoms, while cubes comprised $n = \{14, 63, 172, 365, 666, 1099, 1688\}$ atoms. Measuring from the major axis, these octahedra spanned sizes between 0.8 and 4.4 nm, while cubes (measured by edge length) ranged from 0.4 to 3.2 nm. When computing all ground and excited state properties, the highest angular momentum states in the basis set involves $d$ orbitals. This is important for Al optical property results due to interband transitions; \textit{vide infra}).\cite{Gerard_2015, Douglas-Gallardo_2017}

\subsection{Real-Time Dynamics}
For all calculations, the single particle electronic density matrix $\rho$ was propagated using a leapfrog scheme according to the Liouville-von-Neumann equation of motion;\cite{Bonafe_2020} that is, the first term in
\begin{equation}
	\label{eq1} 
	\frac{\partial\rho}{\partial t} = -i(\mathbb{S}^{-1}\mathbb{H}\rho - \rho \mathbb{H}\mathbb{S}^{-1})
	-(\mathbb{S}^{-1}\mathbb{D}\rho+\rho\mathbb{D}^{\dagger}\mathbb{S}^{-1})\, ,
\end{equation}
where $\mathbb{H}$ is the Hamiltonian matrix and $\mathbb{S}$ is the orbital overlap matrix. The second term accounts for dissipation and allows for energy exchange between electrons and nuclei at the Ehrenfest level of theory. In particular, $\mathbb{D} = \langle\phi_\mu |\frac{\partial\phi_\nu}{\partial t} \rangle$ is the nonadiabatic coupling matrix between localized atomic orbitals $\phi_\mu$ and $\phi_\nu$; introducing this term introduces a mechanism by which nuclear motion can induce electronic transitions, thereby leading to electronic thermalization. Importantly, the incorporation of derivative coupling in equation (1) inexorably leads to photodissociation at sufficient field strengths.\cite{Wu2020} When calculating absorption spectra and dissociation trajectories, we disabled nuclear dynamics for these otherwise thermodynamically unfavorable nanostructures (at these sizes\cite{Burda_2005}); however, we restricted this nuclear dynamics term to \ch{H2} when simulating photodissociation trajectories.

To drive the system, a time-dependent external field $\mathbf{E}(t)$ was added to the DFTB Hamiltonian, where the external potential $V_{A}^{ext}$ was calculated under the electric dipole approximation:
\begin{equation}
	V_{A}^{ext} =- \mathbf{\mu_{A}}\cdot \mathbf{E}(t) = \Delta q_A(t) \mathbf{R}_A(t)\cdot\mathbf{E}(t).
	\label{eq2}
\end{equation}
In equation (\ref{eq2}) the induced dipole $\mathbf{\mu}$ was obtained by summing over the partial charges per atom $\Delta q_{A}$ at nuclear coordinate $\mathbf{R}_A$. Both $\mathbb{D}$ and $V_{A}^{ext}$ couple electronic states to nuclear motion \textit{via} vibronic coupling and field-matter coupling mechanisms, respectively. Subsequently, the force experienced by each atom was computed by averaging over the distribution of electronic states (with total mass $m$) per the Hellmann-Feynman theorem:
\begin{equation}
	m\frac{\partial^{2}\mathbf{R}}{\partial t^{2}} = -\nabla_R\bigg[\frac{\text{Tr}(\rho(t)\mathbb{H}(t))}{\text{Tr}(\rho(t))}\bigg].
\end{equation}
Atomic charges were then calculated by taking the trace over the orbitals centered on atom $A$ per the Mulliken approximation, i.e., $q_{A}(t) = \text{Tr}(\rho(t)\mathbb{S})$.

\subsection{Calculation of Absorption Spectra}
Ground state electronic absorption spectra were calculated following a literature-reported procedure.\cite{Yabana_1996} Briefly, the ground state density matrix $\rho_0$ was "kicked" by a Dirac-delta impulse with strength $\mathcal{E}_0 = 1\times 10^{-3}$\,V/\AA\, of the form $\mathbf{E}_{ind} = \mathcal{E}_0\,\delta (t)\,\mathbf{\hat{e}}$ polarized in direction $\mathbf{\hat{e}}\in \{\hat{i},\hat{j},\hat{k}\}$, thereby exciting all dipole-allowed transitions. Immediately following this perturbation, for a non-orthogonal basis set, the initial density matrix is constructed (per equation (\ref{eq1})) such that
\begin{align}
	\label{eq4}
	\rho(t=0^{+}) = \frac{1}{2}(e^{i/\hbar\hat{V}}&{}\rho_{0}\mathbb{S}e^{-i/\hbar\hat{V}}\mathbb{S}^{-1}\\
	&+\mathbb{S}^{-1}e^{i/\hbar\hat{V}}\mathbb{S}\rho_0 e^{-i/\hbar\hat{V}}),\nonumber
\end{align}
where $\hat{V}$ is the operator equivalent of equation (\ref{eq2}); i.e., $\hat{V} = -\hat{\mathbf{\mu}}\cdot\mathbf{E}_{ind}$ and $\rho_0$ is the ground state density matrix. The nanoparticles' density matrix was allowed to evolve for 50 fs with a timestep of $5 \times 10^{-3}$ fs. The time-dependent dipole moment $\mu_{ij}(t)$, which contains information about the excited frequencies and oscillator strengths, was then obtained by calculating the expectation value of the dipole moment operator ($\mu_{ij}(t) = \text{Tr}(\hat{\mu}\rho(t))$).

Subsequently Fourier transforming the induced dipole determines the frequency-dependent polarizability tensor $\overset\leftrightarrow{\alpha}_{ij}(\omega)$, where
\begin{equation}
	\overset\leftrightarrow{\alpha}_{ij}(\omega) = \frac{\mathbf{\mu}_{ij}(\omega)-\mu_{ij,0}}{\mathbf{E}_{ind}}\, . 
	\label{eq5}
\end{equation}
In practice, the Fourier transform is often damped by a certain time constant (in this case, 5 fs) to level out any "ringing" in the spectra post-processing. Moreover, empirically damping the Fourier transform indirectly accounts for finite lifetime effects in plasmon excitation, which may include radiative relaxation alongside plasmon dephasing and additional scattering processes. \ch{Al} is particularly known for its short plasmon lifetime,\cite{Zeman_1987, Ross_2015, Fonseca_2021} necessitating a small dephasing/damping time; in this work, we damped the Fourier transform by 5 fs. Nonetheless, the damping time used herein is slightly larger than those determined from prior calculations for spherical\cite{Fonseca_2021} and icosahedral\cite{Douglas-Gallardo_2017} \ch{Al} particles due to \ch{Al}'s plasmon sensitivity to shape anisotropy (\emph{vide infra}).\cite{Ross_2015} The absorption cross-section $\sigma(\omega)$ for randomly oriented media was then computed by taking the imaginary component of the trace of the polarizability tensor; that is,
\begin{equation}
	\sigma(\omega) = \frac{4\pi\omega}{3c}\text{Im}[\text{Tr}(\overset\leftrightarrow{\alpha}_{ij}(\omega))]\, ,
\end{equation}
where $c$ is the speed of light.
\subsection{Calculation of \ch{H2} Dissociation Trajectories} 
Laser pulse-driven \ch{H2} dissociation dynamics were computed using a previously reported method.\cite{Giri_2023} An \ch{H2} molecule was vertically centered 2 \AA\, from either the particle vertex or face to determine plasmon localization effects on dissociation probability. The laser pulse was polarized along the bond direction of the \ch{H2} molecule; that is, if the \ch{H2} molecule was situated on the particle vertex, along the $\langle 100 \rangle$ and $\langle 111 \rangle$ directions for octahedra and cubes, respectively. A Gaussian envelope function was then applied to the continuous-wave form of $\mathbf{E}(t)$ in equation (\ref{eq2}), such that
\begin{equation}
	\mathbf{E}(t) = \mathcal{E}_0 e^{-(t-t_c)^{2}/\beta^{2}}\sin(\omega_0(t-t_c))\, ,
\end{equation}  
with a pulse center $t_c$ of 0 fs and driving frequency $\omega_0$ corresponding to the particles' resonant energies obtained from their absorption spectra. The pulse linewidth $\beta = \tau/2\sqrt{\pi}$, was taken to have a pulse duration $\tau$ of 25 fs. However, we propagate the density matrix for at least 50 fs to observe any dynamics post-excitation, such as charge transfer and \ch{H2} dissociation.

 Al octahedra and cubes comprising up to 344 and 365 atoms, respectively, were studied to observe any size dependence on the \ch{H2} dissociation threshold intensity. For each nuclear trajectory, we consider different initial \ch{H2} vibrational coordinates and momenta sampled from the Maxwell-Boltzmann distribution (with $T$ = 300 K). At least 10 trajectories were computed per laser intensity to obtain dissociation statistics. The \ch{H2} molecule was defined to have dissociated if the H-H bond distance exceeded 3 \AA. This procedure was iteratively repeated while decreasing the laser intensity until the dissociation threshold was determined. The laser intensity threshold for \ch{H2} dissociation was defined by the corresponding $\mathcal{E}_0$ below which no dissociation events occurred.
 
 From an experimental standpoint, aluminum's native oxide layer serves as a catalyst support, where van der Waals interactions along with the high porosities typically observed\cite{Koo_2024} increases the mean residence time necessary for electron transfer to induce \ch{H2} dissociation.\cite{Zhou_2016} The incorporation thereof in these calculations, however, not only misrepresents the underlying chemistry of these systems (at these sizes, Al cannot sustain a native oxide layer\cite{Riegsinger_2022}) but also conflicts with the underlying physics of this theory -- the presence of this dielectric layer would further attenuate the short-ranged electronic interactions inherent to this tight binding model.\cite{Douglas-Gallardo_2017} We consequently chose not to incorporate an oxide layer in this work.
 
\section{Results and Discussion}
DFTB is a simplification of Kohn-Sham DFT to a tight-binding form. The most recent iteration of this method involves a third-order Taylor expansion of the Kohn-Sham Hamiltonian around a reference electron density, where perturbations corresponding to charge density fluctuations determine the ground state energy.\cite{Hourahine_2020} Within this framework, the Hamiltonian matrix elements are constructed using pretabulated Slater-Koster combination rules per pairwise element interaction, and exchange-correlation effects (amongst other contributions) are parametrized from higher-level DFT methods within a specified cutoff radius. 

Importantly, due to the many approximations involved and neglect of integral calculations during its computational runtime, DFTB offers the capability to simulate larger systems and electron dynamics over longer time scales. We consequently hypothesized that as \ch{Al} has a prohibitively large (and size-extensive) density of states restricting typical frequency domain calculations, real-time, time-dependent DFTB (RT-TDDFTB) would provide an excellent level of theory to study trends in their plasmonic properties at reasonably large length and time scales otherwise unattainable using typical Kohn-Sham methods.
\subsection{Al Nanocrystals have Broad LSPRs in the UV-vis}
To test this hypothesis, octahedral and cubic \ch{Al} nanocrystals of varying sizes (c.a. 0.5 to 4.5 nm) were constructed and their absorption spectra were calculated according to equation (\ref{eq4}). In the weak-field regime of 0.001 V/\AA, the real-time absorption spectra for small Al particles agree well with those calculated via linear response calculations \textbf{(Fig. S1)}. Octahedra show resonant transitions in the UV that increase in intensity and redshift in resonant energy $E_{res}$ with particle size \textbf{(Fig 1a, b)}. Interestingly, a greater than 0.1 eV shift is observed between \ch{Al19} and \ch{Al44}. Shrinking down a bulk metal into sub-nanoscale dimensions discretizes their otherwise continuous energy levels. The Kubo gap $\delta$, or the average spacing between electronic energy levels, provides a metric to quantify this effect, which is related to the Fermi energy $\varepsilon_F$ and the valence electron number $N$ such that\cite{Kubo1962, Kawabata1966}
\begin{equation}
	\delta = \frac{4 \varepsilon_F}{3N}\,.
\end{equation} 
For Al ($\varepsilon_F = 11.6$ eV, measured relative to the conduction band minimum), only 60 valence electrons are necessary to render the Kubo gap comparable to $k_B T$ at room temperature; that is, only 20 atoms are necessary for thermal fluctuations to induce electronic transitions. Above this, an Al cluster can be considered sufficiently metallic. 
\begin{figure}
	\includegraphics[width=\linewidth]{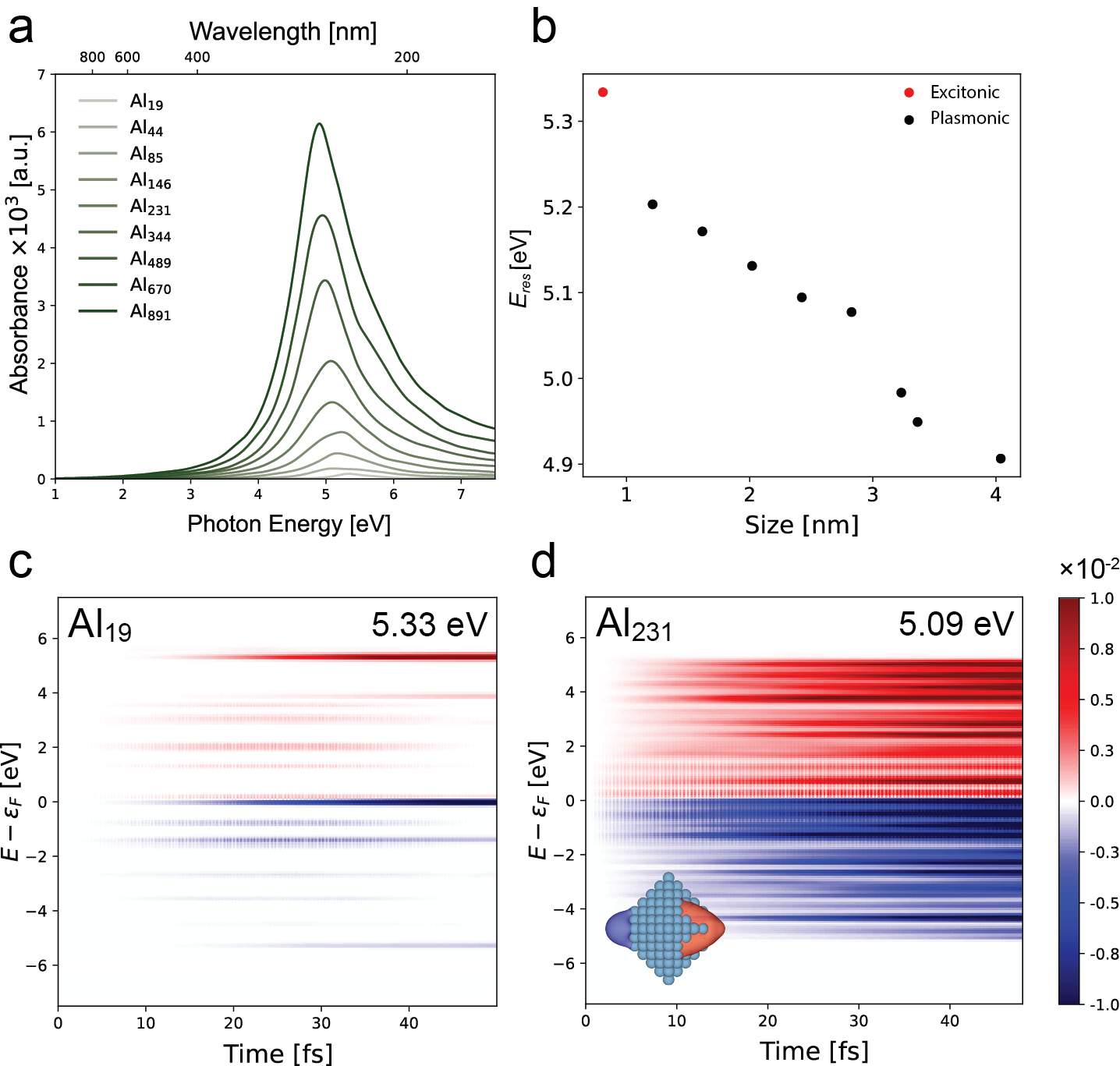}
	\caption{Octahedral Al nanocrystals have LSPRs in the UV. \textbf{(a)} Absorption spectra with particle size. \textbf{(b)} Resonant energy ($E_{res}$) with particle size and resonant mode classification (i.e., excitonic \emph{vs.} plasmonic). \textbf{(c)} Orbital populations as a function of energy $E-\varepsilon_F$ and time for \ch{Al19} and \textbf{(d)} \ch{Al231}. The inset for (d) denotes the charge distribution at the resonant energy, with red being positive, and blue negative.}
\end{figure}
To investigate the nature of this transition, the particles were irradiated with continuous-wave (CW) excitation at $E_{res}$. The orbital populations were then calculated by projecting the molecular orbital populations onto the ground state molecular orbitals. From this, the discrete electronic (de)population with time in \ch{Al19} suggests excitonic character to this resonant transition \textbf{(Fig. 1b)}. In contrast, the orbital populations for \ch{Al231} are roughly continuous, indicating that this resonance is sufficiently collective to be considered plasmonic \textbf{(Fig. 1d)}. At these sizes, the plasmon resonance is dipolar in character and remains so as the particles grow.

 \begin{figure*}[ht!]
 	\centering
	\includegraphics[width=\textwidth]{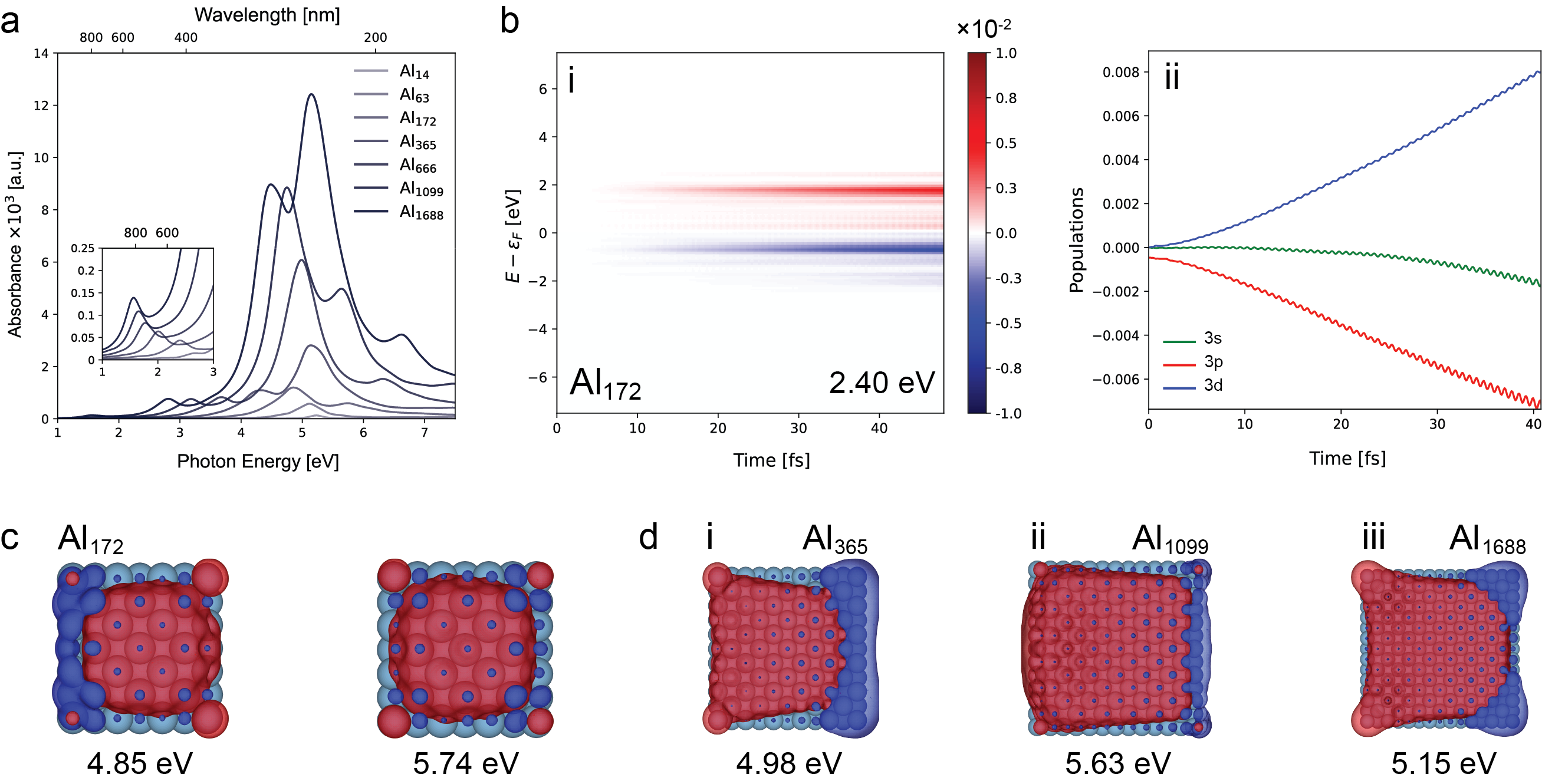}
	\caption{Cubic Al nanocrystals display more complex resonant modes. \textbf{(a)} Absorption spectra with particle size. The inset indicates the region between 1-3 eV containing Al's interband transition. \textbf{(b)} Orbital population dynamics at the interband transition (2.40 eV) for \ch{Al172} by \textbf{(i)} energy, and \textbf{(ii)} shell. \textbf{(c)} Charge density distribution for higher energy modes for \ch{Al172}, corresponding to symmetric (4.85 eV) and antisymmetric (5.74 eV) hybridized modes between the dipolar plasmon resonance with localized corner states. \textbf{(d)} Resonant plasmon modes for Al nanocubes, with varying symmetries.}
\end{figure*}

Cubes, on the other hand, show more elaborate spectra \textbf{(Fig. 2a)}. With increasing particle size, each transition becomes more continuous in nature, commensurate with an increased density of states with more Al atoms per cube \textbf{(Fig. S2)}. Furthermore, these cubes have significantly higher oscillator strengths compared to octahedra (roughly twice as high). The lowest energy transitions likely correspond to Al's spectrally-localized interband transition (with either s$\rightarrow{}$p, or p$\rightarrow{}$d contributions), as evidenced by their discrete nature \textbf{(Fig. 2b)}. Unlike the coinage metals, bulk Al's band structure is such that a near parabolic dispersion around the $W$ and $K$ points with transitions corresponding to 1.5 eV effectively localize all of its interband excitations within this region, thereby enabling strong LSPR responses throughout the optical spectrum without interband damping (\textbf{Fig. S3}).\cite{Ehrenreich_1963} At the same time, quantum size effects redshift the excitation energy from roughly 2.7 eV until it converges to its bulk value. 

The higher energy mode blue of the interband transition corresponds to localized corner states that are hybridized with plasmon modes of the Al cluster. Due to their inability to maximize nearest-neighbor bonds, corner atoms have an increased localized hole density when optically excited.\cite{Kottmann_2001} For \ch{Al172} in particular, the corner-plasmon mode at 4.85 eV shows intense absorption corresponding to oscillations of the electron density that shows mostly dipolar character and where the corner atoms are mostly positive but with oscillations that are in phase with charges on the left and right edges of the cluster.  Meanwhile the darker mode at 5.74 eV is more quadrupolar in nature, with the corner atoms out of phase with charges on the edges of the cube \textbf{(Fig. 2c)}. Due to modal interference, this hybrid dipolar-corner mode is anomalously redshifted (to 4.85 eV) compared to the trends observed in the dipole $E_{res}$ for smaller particles (< 172 atoms) and \ch{Al365}. Similar modal splitting was observed for cuboctahedral particles with geometries containing only one corner atom per vertex \textbf{(Fig. S4a-b)}. Additionally, the proposed corner mode intensity decreased when corner atoms were removed in \ch{Al172}  \textbf{(Fig. S4c)}, which, together, suggests the spatially localized nature of this transition. While the particles grow, charge separation leads to a dramatic redshift by nearly 3 eV, in turn detuning these corner states from the plasmon modes. 

Beyond \ch{Al172}, the dipolar plasmon mode significantly increases in intensity and redshifts due to depolarization effects. Additional higher energy features begin to appear (starting at 6.31 eV for \ch{Al666}) which redshift and significantly increase in intensity more than the dipole mode. Probing the electron distribution for the 5.63 eV mode in \ch{Al1099} shows dipolar character with electron and hole localization at the corners as well, indicating a superposition of dipole and quadrupole modes at this energy. At \ch{Al1688} this feature gains more quadrupolar character at 5.15 eV, while nonetheless maintaining the odd symmetry characteristic of a dipolar mode. \textbf{(Fig. 2diii)} From visual inspection, the lower-energy dipole and the higher-energy hybrid dipole-quadrupole modes for \ch{Al1099} have comparable energies to the split modes observed for \ch{Al172}. Importantly, similar higher-order features are observable for Na cubes (also a free electron metal) of comparable sizes, while analogous features are absent for Ag or Au cubes (which both have strong $d\rightarrow{}sp$ interband contributions; \textbf{Fig. S5}). Fuchs has previously shown that the dielectric susceptibility of ionic crystal cubes with sizes below the quasistatic limit can be expressed as a sum of normal modes that are determined by particle symmetry.\cite{Fuchs_1975} While studied in the context of their infrared absorption spectra, similar arguments can be made for the UV-vis-NIR absorption behavior of cube-shaped metal clusters, with two dominant spectral peaks at lower photon energies (analogous to the two modes we see in the 4.5-6 eV range) and lower-intensity higher order modes at higher energies.\cite{Ruppin_1996} This, then, underscores modal interference alongside a combination of particle geometry and Al's free-electron nature as the predominant contributors to the more elaborate spectral features observed for Al nanocubes.

\subsection{Plasmonic Al Nanocrystals can Induce \ch{H2} Dissociation}
Real-time dynamics offers a powerful approach to simulate the photocatalytic response of plasmonic systems by directly propagating the single-particle density matrix under an explicit time-dependent electric field. We then hypothesized that, as RT-TDDFTB is able to reproduce Al's plasmonic properties with high computational efficiency, the use of this theory would enable ensemble-based trajectory sampling to analyze photocatalytic reactions of interest. As a model reaction, we chose the dissociation of \ch{H2}. This is perhaps one of the simplest reactions, yet one of the most meaningful -- catalytic hydrogenation is one of the largest branches of industrial heterogeneous catalysis. However, \ch{H2} forms a strong, nonpolarizable $\sigma$ bond with a dissociation enthalpy of 4/5 eV, a barrier insurmountable by thermal activation alone.\cite{Mukherjee_2013} Hot carriers are well-known to catalyze this reaction;\cite{Mukherjee_2013, Giri_2023, Li_2023} the use of DFTB in particular would allow for a statistically meaningful analysis of this process. 

To test this hypothesis, we simulated at least 10 different trajectories per particle size and irradiation intensity explored for both cubes and octahedra when subject to a Gaussian pulsed excitation at the LSPR frequency. RT-TDDFTB calculations do not incorporate electron-electron and electron-phonon scattering, preventing them from fully capturing the electronic relaxation processes that should dominate at longer timescales.\cite{Bonafe_2020, DouglasGallardo_2021, Giri_2023} Consequently, we propagated the density matrix for 50 fs (\textbf{Fig. 3}), with the understanding that the metal particle will be energetically hot at the end of the calculation, thereby enabling sufficient charge transfer to the nearby \ch{H2} molecule to induce dissociation. The time-dependent dissociation probability (calculated from each ensemble of trajectories per irradiation intensity) was determined \textit{via} Gaussian binning (centered around 3 \AA\, with a width of 1.3 \AA) along interatomic distances ($d_{\ch{H}-\ch{H}}$). Bond distances at the equilibrium \ch{H}-\ch{H} bond length of 0.75 \AA\, were assumed to have a dissociation probability of zero and as soon as the interatomic distance passed the center of the Gaussian (3 \AA) the probability was maintained at its peak value of 1. Dissociation strongly depends on the initially sampled vibrational configuration, as is seen in the trajectories. For \ch{Al344H2}, at low intensities, $d_{\ch{H}-\ch{H}}$ shows coherent oscillations around the equilibrium bond length without dissociation. At the threshold intensity of $2.12 \times 10^{12} \text{ W/cm}^{2}$, 80\% of the trajectories lead to dissociation. Above this, all of the trajectories are dissociative (save for one at $3.32 \times 10^{12} \text{ W/cm}^{2}$ with a strong oscillation amplitude), with the dissociation probability saturating at a finite value within the first 25 fs (i.e., during interactions with the pulse).

\begin{figure}[t]
	\includegraphics[width=\linewidth]{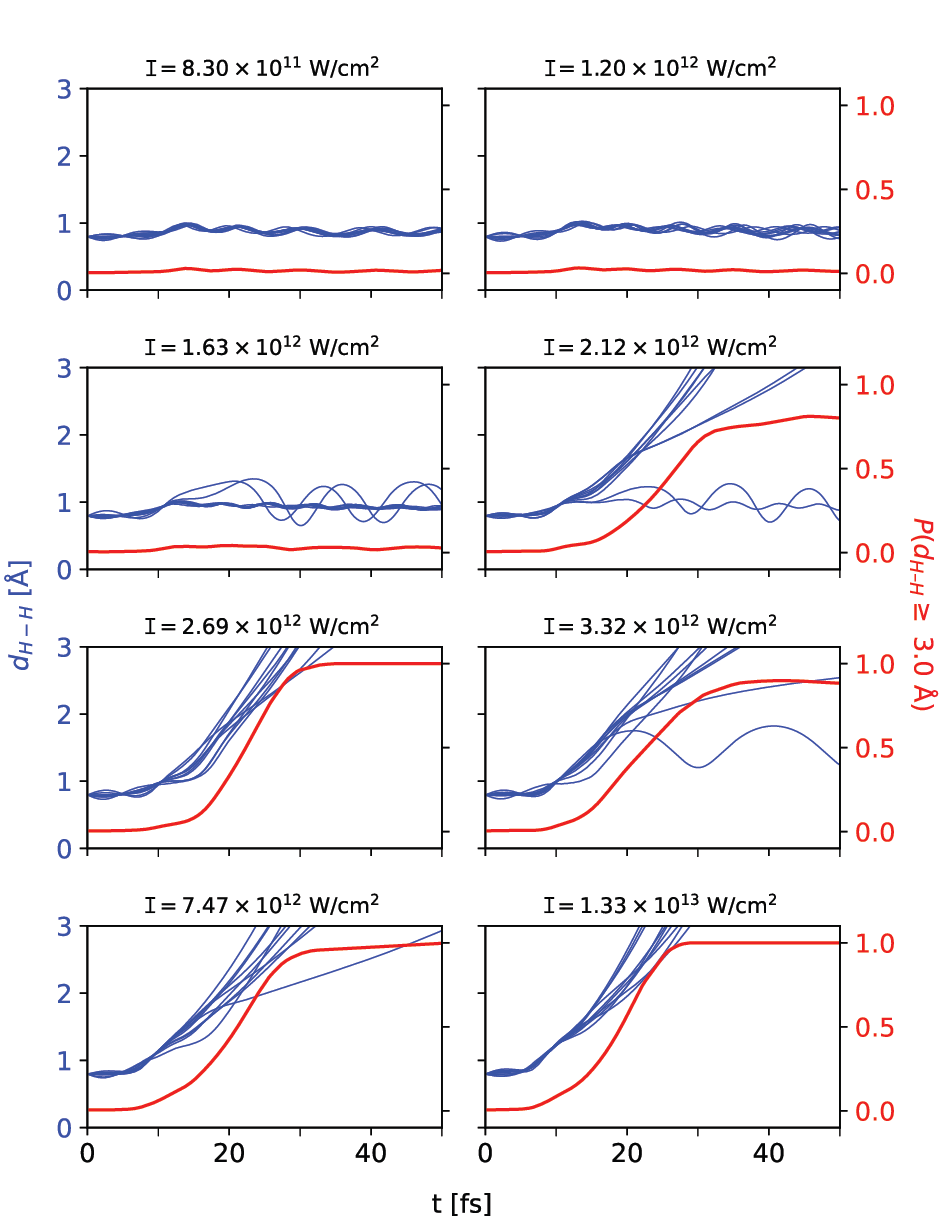}
	\caption{Configurational sampling from the Maxwell-Boltzmann distribution yields an ensemble of \ch{H2} dissociative trajectories. \ch{H}-\ch{H} bond distances ($y_1$ axis) are shown with respect to time for \ch{Al344H2} as the electric field strength $\mathcal{E}_0$ (and therefore irradiation intensity $I$) is gradually increased at its plasmon energy of 5.08 eV. The $y_2$ axis analyzes the probability of \ch{H2} dissociation \textit{via} Gaussian binning along interatomic distances at each intensity.}
\end{figure}

To gain an understanding of the electron dynamics involved during \ch{H2} dissociation, we examined the electronic response following particle irradiation at the threshold intensity for \ch{Al344H2} (\textbf{Fig. 3}). Notably, many of the dissociative trajectories examined herein show net electron transfer within the first 12.2 fs to H$_2$ followed by dissociation in which \ch{H-} binds to the cluster. This is apparent in Fig. 4b, where the charge oscillations move to more positive values (measured in units of electron charge) at a time when over 5 eV of energy has been transferred to the molecule, then the electron transfer is reversed, leading to \ch{H-} on the surface. Due to impedance mismatch, Al's phonons only weakly couple to \ch{H2}'s high frequency vibrational mode. Consequently, \ch{H2} is known to not readily adsorb onto Al surfaces under ambient conditions.\cite{Hammer_1992, Paul_1988, Hammer_1993, Henry_2009, Wolverton_2004} We thus attribute the dynamics that we see to weak Coulombic attraction between the positively-charged Al cluster and the hydride anion after the electron transfer is reversed. 

The time-dependent molecular charge $\Delta Q$ was further examined for a nondissociative trajectory (\textbf{Fig. 4b}). Following the pulsed excitation, electron exchange leads to the dipole intensity strongly oscillating between the two subsystems with a frequency corresponding to that of the LSPR. Around 15 fs, the oscillatory charge "sloshing" is more symmetrical about zero than for the dissociative trajectory, and the net energy transfer is smaller. As a result there is little backtransfer and dissociation does not occur. Overall the dissociative trajectory achieves a higher maximum in $\Delta E_{H2}$ compared to the nondissociative one, which, in turn, dictates the different outcomes.

\begin{figure*}[t]
	\centering
	\includegraphics[width=\textwidth]{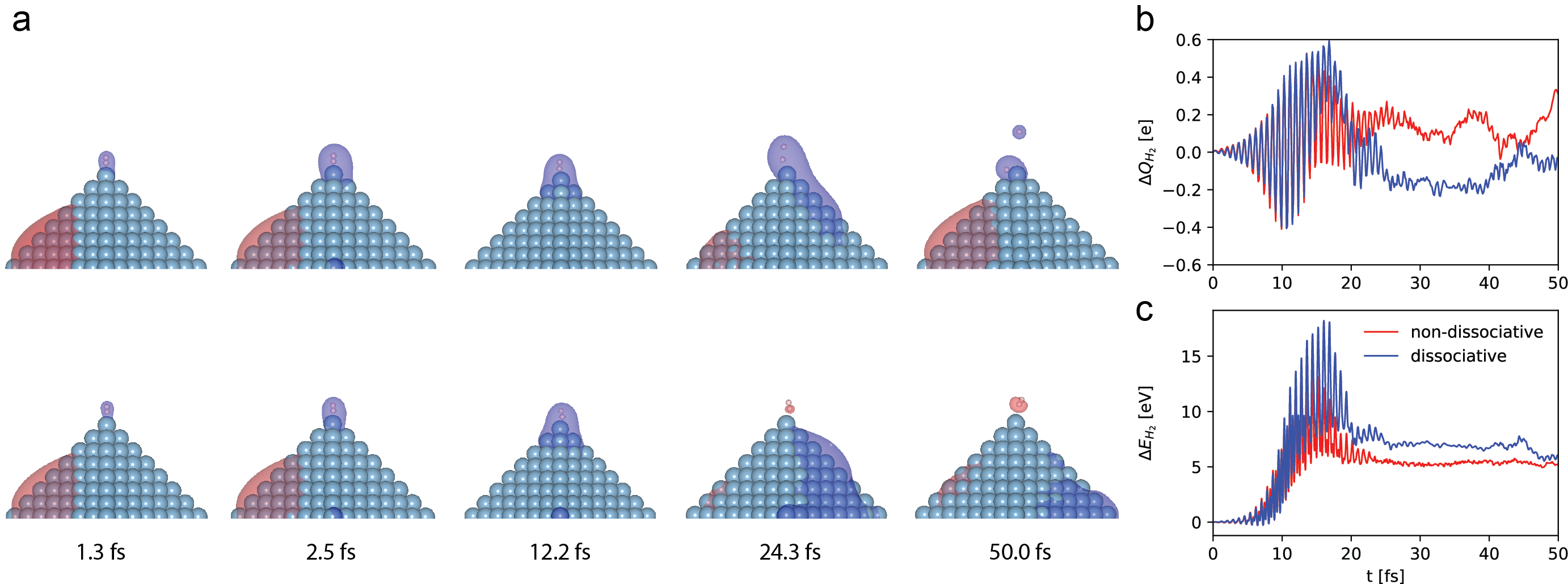}
	\caption{Electron transfer occurs midway through the pulsed interaction. \textbf{(a)} Charge density differences (per the Mulliken approximation) plotted with time. \textbf{(b)} \ch{H2} molecular charge plotted as a function of time for dissociative (blue) \textit{vs.} non-dissociative (red) trajectories. An octahedral \ch{Al344H2} nanocrystal was irradiated at its threshold intensity of $2.12 \times 10^{12}$ $\text{W/cm}^{2}$.}
\end{figure*}

Further interrogation of the electron-hole pair (EHP) distribution under pulsed excitation shows the transient production of hot electrons (\textbf{Fig. 5}). Following plasmon excitation, EHPs are produced with energies approximately $\pm$ 1 eV relative to $\varepsilon_F$ (\textbf{Fig. 5a}). At around 12 fs, the EHP distribution broadens with the largest concentration of EHPs produced in the vicinity of the LSPR energy. Thereafter, continuous photon absorption and scattering events during the pulse leads to energy redistribution and electronic thermalization with a fraction of electrons having energies up to 20 eV above $\varepsilon_F$, while the hole distribution remains within 5 eV of $\varepsilon_F$. This particular EHP distribution can be connected to Al's band structure: weak valence band dispersion around the $X$ and $W$ high-symmetry points neighboring $\varepsilon_F$ enable efficient electron excitation to a high density of conduction band states (\textbf{Fig. S3}). The lack of accessible valence bands significantly below $\varepsilon_F$ at the $X$ point limits the depth of hole states that can be generated during excitation. As a result, electron excitations are restricted in energy, though the high density of conduction band states enables continuous excitation to higher energy states well above the LSPR energy. As longer-time effects from electron-electron and electron-phonon scattering are not fully captured at this level of theory, the distribution converges beyond 24.3 fs. 

\begin{figure}[ht!]
 	\centering
	\includegraphics[width=\linewidth]{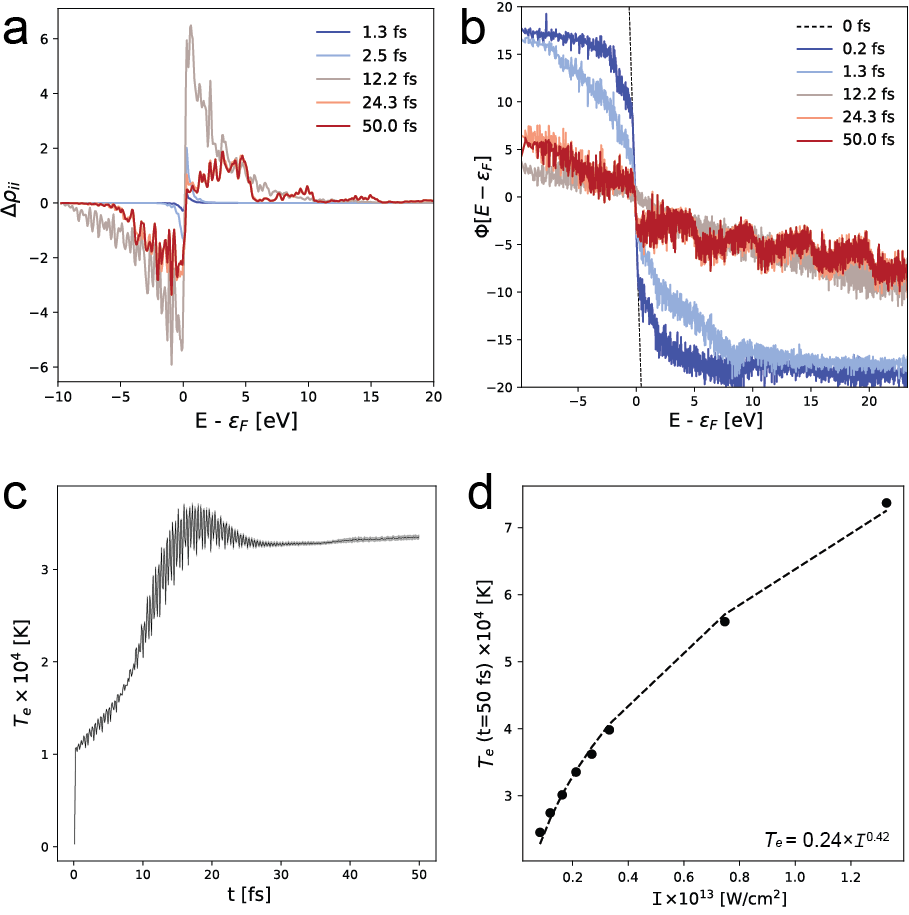}
	\caption{Aluminum rapidly thermalizes. \textbf{(a)} Transient electron energy distributions, calculated from the difference in orbital occupations relative to $t = 0$ fs. \textbf{(b)} Quasilogarithmic representation of energy distributions where $\phi[E,t] = -\ln(1/f(E,t)-1)$ and $f(E,t)$ is the electron energy distribution at time $t$. \textbf{(c)} Mean electronic temperature with time sampled over ten different trajectories. Standard deviations are shaded grey. \textbf{(d)} Mean electronic temperature dependence with pump intensity $I$. Temperatures were taken at $t = 50$ fs with a power law fit shown in the bottom right. At least ten trajectories were performed at each irradiation intensity. All trajectories were performed for \ch{Al344H2} and irradiated at the plasmon energy of 5.08 eV.}
\end{figure}

To obtain the effective translational temperatures associated with each time, the nonequilibrium electron distributions for Al$_{344}$H$_2$ were fit to a quasilogarithmic distribution along each trajectory (\textbf{Figs. 5b-c}). In this representation, an equilibrium Fermi-Dirac distribution appears as a straight line with its slope inversely proportional to the electronic temperature. We find that the nonequilibrium electronic energy distributions show the general features of ultrafast electron dynamics in metals, akin to those reported in previous Boltzmann transport simulations.\cite{Rethfeld_2002,Mueller_2013} Immediately following the pulse, the equilibrium electronic temperatures $T_e$ corresponding to 300 K are driven far from thermal equilibrium to temperatures over 2 orders of magnitude higher while absorbing over 800 eV of energy (2.3 eV per atom) during the pulse (\textbf{Fig. S7}). \textbf{Fig. 5a} shows large changes in the populations for resonant excitation (5.08 eV), and then the quasilogarithmic representation shows considerable broadening due to the large increase in $T_{e}$, with a smooth distribution at 12.2 fs that evolves at 24.3 eV to a series of 5 eV step structures. Note that the step structure is already in place before the pulse is over, which suggests that it is a result of the excitation dynamics itself rather than subsequent scattering processes. Off-resonant excitation (not close to the plasmon resonance) preserves the step-like features observed, with the step width still commensurate with the excitation frequency (now 3 eV; \textbf{Fig. S8}). From this, the step structure is clearly associated with the driving frequency rather than the intrinsic band or energy level structure associated with the Al cluster. We infer that these features arise from  excitations up from levels that are integer multiples of the incident photon energy. These can occur starting from levels at the Fermi level, and down to $\hbar \omega$ below the Fermi level, through resonant excitation. Then there is a noticeable reduction in intensity as one goes to higher excitation energies as this requires multiphoton transitions. This leads to steps above the Fermi level, and moreover this also can generate steps below the Fermi level at sufficiently high intensities. We also see steps at a subthreshold intensity of $6.6 \times 10^{10}$ W/cm$^2$, though with a more pronounced step function-like behavior below $\varepsilon_F$ and the total energy absorbed is over three orders of magnitude lower than that at the threshold (\textbf{Fig. S9}). The mean $T_e$ peaks at nearly 37,000 K around 18 fs, before decreasing and stabilizing at approximately 33,000 K for the remaining 25 fs (\textbf{Fig. 5c}). This maximum in $T_{e}$ closely follows the trends discussed in \textbf{Fig. 4} as well as the inflection point in the total absorbed energy (\textbf{Fig. S7a}). However, inefficient relaxation pathways in our RT-TDDFTB model preclude dissipation that would wash out the step structure and also reduce $T_e$ at longer times, as would be expected from two-temperature empirical models that are often used to describe the time dependence of electron populations.\cite{Carpene_2006, Hashimoto_2012}

Generally, the mean final electronic temperature (at 50 fs) exhibits a characteristic power law dependence with irradiation intensity (\textbf{Fig. 5d}). Typically, the exponent in this scaling law reflects the material's ability to absorb and redistribute energy.\cite{Mora_1982, Kluge_2011, Hashimoto_2012} The exponent from this fit of 0.42 is roughly similar to those predicted by ponderomotive scaling laws at higher intensities (> $10^{17}$ W/cm$^2$).\cite{Rusby_2024, Chen_2009, Kluge_2011} Together, aluminum's free electron nature allows for efficient energy redistribution as its conduction band is less affected by interband transitions compared to other metals. This enables the rapid thermalization observed here, contrasting with metals like gold or copper, where strong interband coupling leads to slower energy relaxation.\cite{Del_Fatti_1998, Del_Fatti_2000, Rethfeld_2002, Arbouet_2003, Hashimoto_2012} Furthermore, the neglect of photothermal effects in this theory precludes their role in \ch{H2} dissociation, underscoring rapid energy transfer from hot electrons as the predominant contributor to \ch{H2} dissociation on Al nanocrystals.

 Further, we find that the dissociation threshold intensity $I_{thresh}$ decreases with particle size for both octahedra and cubes, following a power law trend comparable to that found previously for Au and Ag (\textbf{Fig. 6a}).\cite{Giri_2023} This decrease is more pronounced for octahedra than for cubes, likely due to the relative contributions of the curved regions compared to their flat faces with particle size. Indeed, the sharp corners of cubes create regions of high local curvature that lower the threshold for smaller particles compared to octahedra. However, beyond a critical crossover size of roughly 3.8 nm, the cubes' flat faces (with an infinite radius of curvature) become the dominant factor, while their corner geometry remains unchanged. In contrast, octahedra maintain their curved surfaces at all scales, with the concurrent increase in transition moment allowing them to more effectively concentrate electromagnetic energy into sub-wavelength volumes as they grow larger. We also found that the dissociation threshold is significantly greater for \ch{H2} molecules localized at the particle faces compared to the vertices, consistent with stronger field enhancement (and charge localization) at the vertices (\textbf{Fig. S10}). Another salient point is that the smallest particles examined (\ch{Al19} and \ch{Al14}) deviate from this expected power law behavior (\textbf{Fig. 6a}), attributable to the excitonic (as opposed to plasmonic) character of its primary absorption peak (\textit{vide supra}). Beyond this, $I_{thresh}$ decreases monotonically with particle size, with power law fit parameters provided in \textbf{Table 1}. The final electronic temperature also exhibits a similar trend, with an anomalous increase between \ch{Al19} (\ch{Al14} for cubes) to \ch{Al44} (\ch{Al63}), followed by a monotonic decrease thereafter.

\begin{figure}[ht!]
 	\centering
	\includegraphics[width=0.7\linewidth]{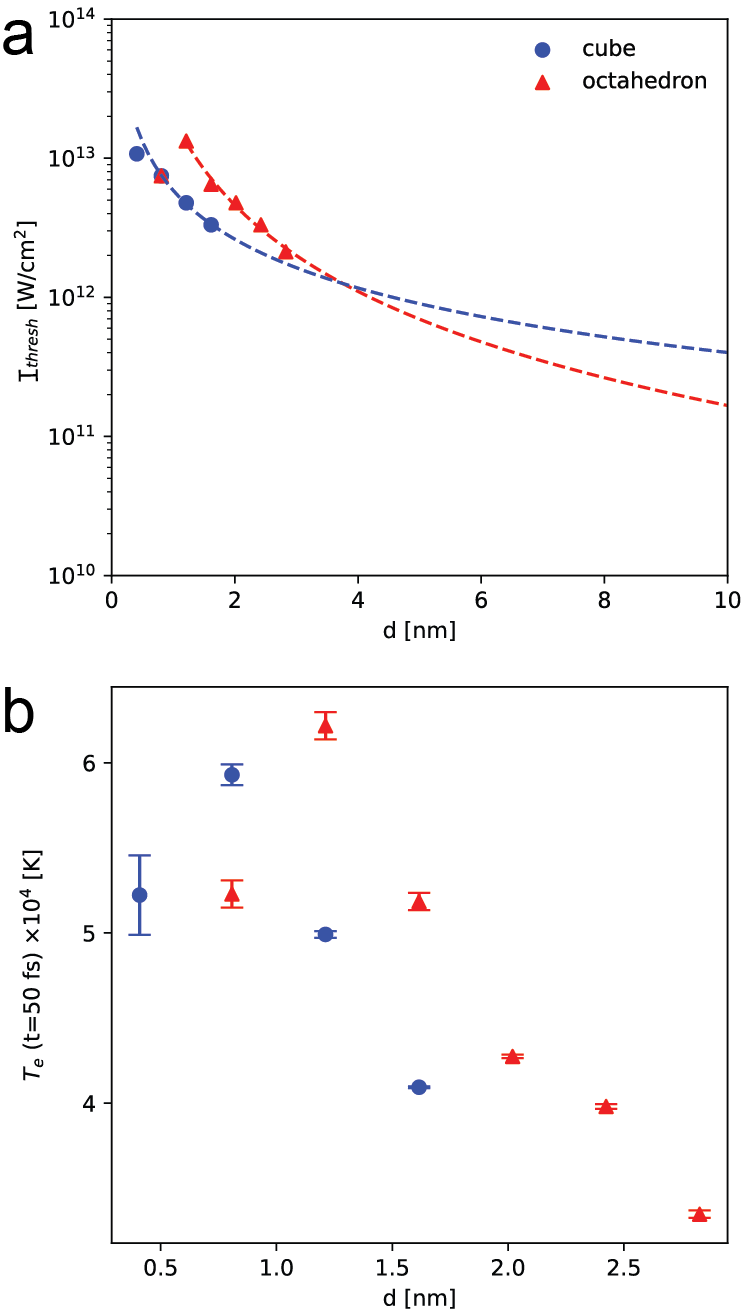}
	\caption{The dissociation threshold intensity generally decreases with larger particles. \textbf{(a)} Threshold intensity for \ch{H2} dissociation with particle size, for octahedra (red triangles) and cubes (blue circles). Trendlines are shown as dotted lines. Due to the multiple modes observed, cubes were irradiated at their principal absorption peak. \textbf{(b)} Mean electronic temperature $T_{e}$ at 50 fs with particle size, irradiated at the corresponding $I_{thresh}$ for each particle size.}
\end{figure}

\begin{table}[h!]
    \centering
    \begin{tabular}{l|l|l|}
        & $\alpha$ & $\beta$ \\ 
        \hline
        octahedra & $1.92\times10^{13}$ & $-2.06$ \\
        cube & $5.87\times10^{12}$ & $-1.17$
    \end{tabular}
    \caption{Power law fits of threshold dissociation intensity with particle size $I_{\text{thresh}} = \alpha\times d^{\beta}$.} 
    \label{tab:power-law}
\end{table}

The trends examined herein indicate that $I_{thresh}$ should significantly decrease when extrapolated to experimentally-used particle sizes. For a 100 nm octahedral nanocrystal, extrapolation would yield a threshold intensity reaching $1.45\times$ 10$^9$ W/cm$^{2}$ -- a decrease by nearly four orders of magnitude. However, this still presents a significant deviation from experimentally-observed intensities (using CW excitation) by approximately six orders of magnitude.\cite{Zhou2016} We have recently discussed some limitations intrinsic to this model that may influence this discrepancy.\cite{Kar_2025} A key factor is the pulse duration. The use of longer pulse durations (as used experimentally, $\sim 10^2$ fs) should confer additional time for charge/energy transfer to the molecule. This gradual energy uptake would also mitigate electronic relaxation due to limitations in describing electron-electron scattering at this level of theory; Ehrenfest dynamics do not fully capture stochastic energy dissipation due to an incomplete many-body treatment. Importantly, the excitation energy is not utilized completely in molecular dissociation (\textit{vide supra}; \textbf{Fig. S7}), with only a fraction of the total absorbed energy contributing to reaction dynamics, while the remainder is redistributed through competing relaxation pathways. Moreover, as only one \ch{H2} molecule is considered per particle, LSPR excitation does not lead to efficient energy transfer in this model. In a more realistic scenario with multiple molecules in the vicinity of the particle (and particle aggregates with higher localized electromagnetic fields), plasmonic hot carriers could interact collectively with the reactant ensemble, potentially enhancing the dissociation efficiency.\cite{Kar_2025}

\begin{figure}[ht!]
 	\centering
	\includegraphics[width=\linewidth]{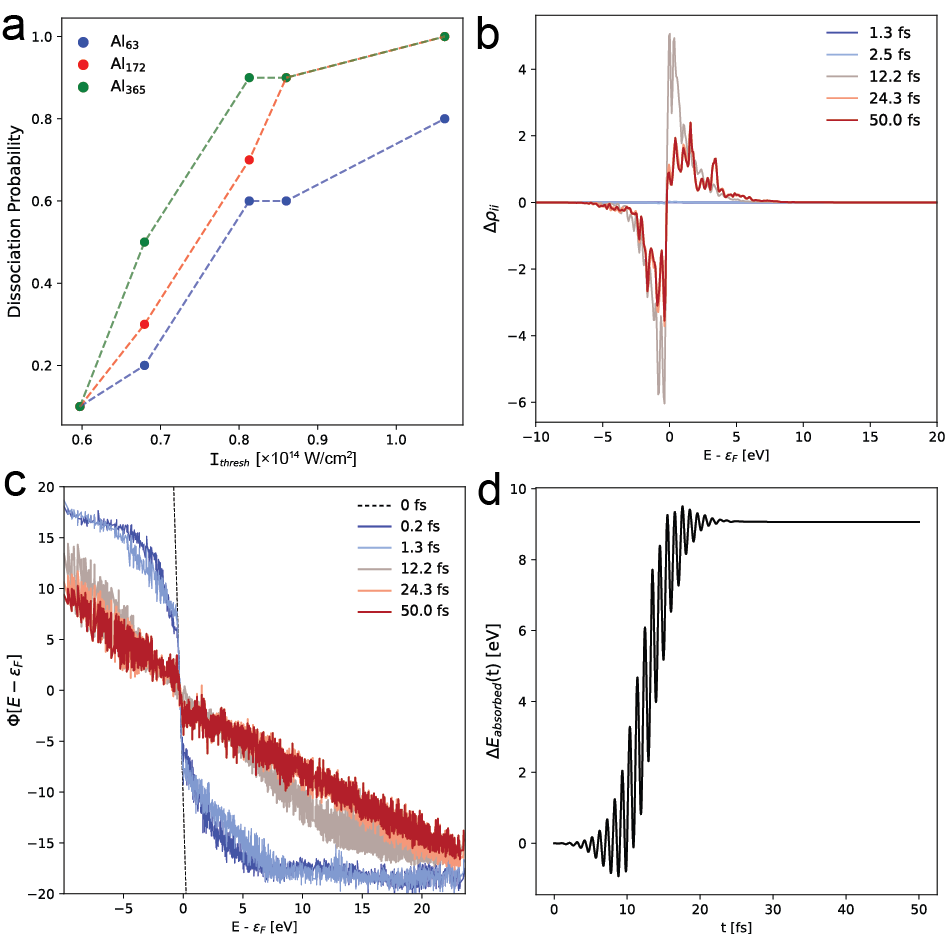}
	\caption{Interband excitation for small Al particles is less efficient in catalyzing \ch{H2} dissociation. \textbf{(a)} Threshold intensity for \ch{H2} dissociation with particle size for Al nanocubes. \textbf{(b)} Transient electron energy distributions, calculated from the difference in orbital occupations relative to $t = 0$ fs. \textbf{(c)} Quasilogarithmic representation of energy distributions at the threshold intensity of $6\times 10^{13}$ W/cm$^{2}$ for \ch{Al365}. \textbf{(d)} Total energy absorbed relative to time $t = 0$ fs for \ch{Al365}. Particles were irradiated at their corresponding interband transitions of 2.67 eV (\ch{Al63}), 2.39 eV (\ch{Al172}), and 2.0 eV (\ch{Al365})}
\end{figure}

Contrary to expectations in the literature,\cite{Swearer_2016, Zhou_2016} we find that exciting the optically-allowed interband transition in Al nanocubes is less efficient in catalyzing \ch{H2} decomposition compared to plasmon irradiation. Al's high density of states and localized interband transition (\textit{vide supra}) should in principle promote efficient carrier generation - and thus \ch{H2} dissociation - when excited. However, our calculations reveal that excitation of this low oscillator strength interband transition requires nearly an order of magnitude higher threshold intensity for dissociation. Although larger particles show an increased dissociation probability with higher pump fluence, the threshold intensity of $6\times 10^{13}$ W/cm$^2$ remains invariant across particle sizes (\textbf{Fig. 7a}). The electronic energy distributions maintain symmetry around $\varepsilon_{F}$ (\textbf{Fig. 7b}), but with notably lower electronic temperatures, as evidenced by the steeper slopes in \textbf{Fig. 7c}. The convergence of total absorbed energy around 9.5 eV (\textbf{Fig. 7d}), in tandem with these observations, altogether suggests that the oscillator strength primarily dictates a transition's effectiveness for \ch{H2} dissociation at these sizes.
 
While this observation ostensibly contrasts with what is reported experimentally,\cite{Swearer_2016, Zhou_2016} these findings can be understood by considering aluminum's unique properties as a UV-active post-transition metal. While larger Al particles exhibit predominantly radiative plasmon decay,\cite{Ross_2015, Fonseca_2021} the small particle sizes studied here ensure purely nonradiative plasmon decay through intrinsic (i.e., Landau) damping. Consequently, the reduced efficiency of interband excitation in this size regime stems from the nonradiative character of the plasmon itself. At these sizes, the oscillator strength dictates the particles' catalytic efficiency. As small (<10 nm) Al particles are theorized to have high absorption efficiencies and are incapable of supporting oxide layers which would otherwise induce chemical interface damping,\cite{Riegsinger_2022} plasmon excitation could lead to high catalytic turnover for many reactions of interest.

\section{Conclusions}
Aluminum is indeed the prototypical free electron metal. Its ability to sustain strong plasmon resonances throughout the UV-vis-NIR confers an exquisite optical tunability. Through a purely quantum level of theory used to describe the optical properties of Al nanoparticles, we report that these effects can be observed even at quasistatic length scales simply by tuning the particle size, shape, and thus ability to support different symmetry-dependent plasmonic normal modes. The high electronic temperatures and energies observed during the pulse itself dictate the outcome of \ch{H2} dissociation, and the efficiency of such is found to scale with particle size. Moreover, irradiation at the plasmon frequency is more efficient than interband excitation for these small sizes due to its predominantly nonradiative character. Taken together, this study has shown that density functional tight-binding can indeed be used to model the photocatalytic performance of reasonably-sized aluminum nanoparticles and that their properties are inherently different from those of the coinage metals. The significant gains in computational runtime at this level of theory not only allow for a deeper understanding of the rich physics in this free-electron metal but also a statistically meaningful analysis of this most abundant metal in the Earth's crust.
\begin{suppinfo}
\end{suppinfo}
\section{Acknowledgements}
N.S.C. is grateful to Dr. Charles Jason Zeman IV for mentorship and Dr. Sajal K. Giri for informative discussions.
\section{Funding}
N.S.C. and G.C.S. acknowledge support by NSF grant no. CHE-2347622. N.S.C. gratefully acknowledges support from the National Science Foundation Graduate Research Fellowship through grant no. DGE-22334667. Any opinions, findings, and conclusions or recommendations expressed in this material are those of the authors and do not necessarily reflect the views of the National Science Foundation. Computational time was provided in part by the Quest High-Performance Computing facility at Northwestern University, which is jointly supported by the Office of the Provost, the Office for Research, and Northwestern University Information Technology.
\bibliography{Al_DFTB.bib}
\end{document}